# A novel shape-based loss function for machine learning-based seminal organ segmentation in medical imaging

Reza Karimzadeh[1], Emad Fatemizadeh[1,†], Hossein Arabi[2]

[1]School of Electrical Engineering, Biomedical Signal and Image Processing Laboratory (BiSIPL), Sharif University of Technology, Tehran, Iran

[2]Division of Nuclear Medicine and Molecular Imaging, Geneva University Hospital, CH-1211 Geneva 4, Switzerland

*Abstract*

Automated medical image segmentation is an essential task to aid/speed up diagnosis and treatment procedures in clinical practices. Deep convolutional neural networks have exhibited promising performance in accurate and automatic seminal segmentation. For segmentation tasks, these methods normally rely on minimizing a cost/loss function that is designed to maximize the overlap between the estimated target and the ground-truth mask delineated by the experts. A simple loss function based on the degrees of overlap (i.e., Dice metric) would not take into account the underlying shape and morphology of the target subject, as well as its realistic/natural variations; therefore, suboptimal segmentation results would be observed in the form of islands of voxels, holes, and unrealistic shapes or deformations. In this light, many studies have been conducted to refine/post-process the segmentation outcome and consider an initial guess as prior knowledge to avoid outliers and/or unrealistic estimations. In this study, a novel shape-based cost function is proposed which encourages/constrains the network to learn/capture the underlying shape features in order to generate a valid/realistic estimation of the target structure. To this end, the Principal Component Analysis (PCA) was performed on a vectorized training dataset to extract eigenvalues and eigenvectors of the target subjects. The key idea was to use the reconstruction weights to discriminate valid outcomes from outliers/erroneous estimations. To evaluate the proposed segmentation framework, a 3D fully convolutional neural network was trained with a conventional loss function (Binary Cross-Entropy) and the proposed loss function. These models were evaluated using two separate datasets for hippocampus and heart segmentation tasks. When the proposed shape-based loss function was employed instead of the Binary Cross-Entropy, the Dice scores improved from $0.81\pm0.03$ and $0.74\pm0.07$ to $0.86\pm0.03$ and $0.87\pm0.05$ for hippocampus and heart segmentation from MR and CT images, respectively. Moreover, no outliers, patchy, and unrealistic estimations were observed when the proposed PCA-based loss function was exploited. The quantitative evaluation of the segmentation results indicated that the proposed PCA-based loss function would lead to significant improvements in organ/structure segmentations from medical images.

## I. INTRODUCTION

The purpose of semantic segmentation is to categorize pixels of an image into particular classes, each of which represents a specific organ/structure in an object. Despite tremendous progress in computer vision, automated segmentation is yet regarded as a challenging task in medical image analysis, natural image processing, and search engines [1-4]. Medical image segmentation plays a critical role in diagnosis, treatment planning, and monitoring of diseases, such as cancer [5-8]. To this end, manual segmentation is used for organ and tumor delineation as a reference; however, this task is highly time-consuming and tedious for experts. Furthermore, there are chances for human errors, as well as large intra- and inter-rater variabilities due to fatigue or workload [9]. To address these challenges, many automated and semi-automated algorithms, such as contour-based (e.g., snake and level-set), fuzzy logical, and atlas-based algorithms have been proposed [10-12].

An excellent medical image segmentation algorithm should be accurate, robust, and immune to outliers [13]. With the revolution of deep Convolutional Neural Networks (CNN) in image processing [14, 15], much attention was attracted to the exploitation of the CNN models for image segmentation, which are tremendously improved, compared to conventional approaches [16]. The CNN-based medical image segmentation methods would be able to provide clinicians with accurate and time-efficient organ/tumor segmentation for an effective diagnosis and treatment [2].

The central cores in the development of deep neural networks are 1) network architecture with sufficient trainable parameters and a high nonlinearity for learning complex relations between inputs and outputs, 2) appropriate cost/loss function to enforce/guide the network to learn the underlying task(s), 3) a training dataset (in terms of size, as well as accuracy), and 4) an optimization algorithm for tuning learnable parameters [17]. Any suboptimal setting/configuration at any of these layers would lead to overall sub-optimal performance by the model. In this light, recent studies have demonstrated that despite the extraordinary capacity of deep learning methods in high-level feature extraction, these methods have shown suboptimal performance to learn the shapes and morphology of the target subjects when loss functions are not properly designed [18].

In deep learning-based models for medical image segmentation, final segmentation outputs may suffer from artifacts, such as fragmentation of integrated structures, islands of voxels, and topological discrepancies due to the suboptimal modeling of the shape features. Several strategies have been adopted to address this shortcoming of deep learning, with the focus on enhancing the capacity of the CNN models to capture shape/morphology features via providing prior knowledge and postprocessing the CNN predictions. These approaches could be categorized into three main classes: 1) Conditional/Markov random fields (MRFs) 2) active shape models (ASM), and 3) active contour models (ACM) [12, 19].

The MRFs theory relies on the contextual information between pixels and high-level features of the input image to encourage realistic shape estimation. In a segmentation problem, the probability of a pixel belonging to a specific class is conditioned to the neighborhood pixels that enable contextual perception [20]. Conditional Random Fields (CRFs), similar to MRFs, rely on statistical models for predicting the target structure in the segmentation and labeling problems [21]. The MRFs/CRFs are either employed as post-processing steps [22-24] or used within end-to-end training of the neural networks [25-27] to improve the final segmentation accuracy.

The ASM tends to capture the underlying shape variations from landmarks of accurately annotated images in a training set. Given the landmarks are defined on the training dataset, Principal Component Analysis (PCA) is used to extract weighted eigenvectors, which enable the generation/prediction of new realistic shapes [28]. The ASM could also be utilized as a post-processing step in CNN-based segmentation frameworks, in which erroneous outputs, such as cases with islands of voxels or holes, are processed and corrected to comply with the genuine shape variations. In this light, post-processing shape compliance approaches could recover the structure of the target subject or prevent gross segmentation errors [29-31]. Moreover, the ASM could be employed in CNN-based segmentation frameworks to provide prior knowledge regarding the underlying valid structural variations of the target subject. In addition, the ASM could produce coarse segmentation, on which the CNN models would rely to better estimate the structure of the target subject [32, 33].

The ACM employs deformable contours which can minimize an energy cost function to fit the boundaries of the target subject. Snake and Level Set Function (LSF) are the most well-known methods of the ACM approach [34, 35]. The ACM, as well as the ASM and CRFs, could be employed together with a CNN-based segmentation framework, wherein the CNN segmentation output is regarded as an initial estimate for the ACM methods [36] or vice versa [37].

These approaches are indeed able to improve final segmentation output via post-processing or preparing an initial guess (prior knowledge); however, the issue of suboptimal CNN-based segmentation models to capture the underlying shape and morphology variations are not yet properly addressed [18]. Some efforts have been made to guide CNN models to learn intrinsic shape features and thus, improve the accuracy of the segmentation procedure through defining novel cost functions which encourage the network to learn the underlying shape features. Mohagheghi et al. [38] proposed a convolutional auto-encoder to learn the shape characteristics of the liver. These features were then employed to form a hybrid loss function in combination with the Dice score. Ravishankar et al. [39] exploited two cascade CNNs, wherein the first one predicts the primary mask, and then a shape regularization network performs the final refinement given the ground-truth mask in the training stage. A custom cost function is then defined to train these cascade networks, which encourages the networks to predict the target shape close to the ground-truth shape space.

Definition of a shape-based cost/loss function is an efficient way to guide CNNs to learn the underlying shape variation of the subject. Yet, most attempts in this field use an auxiliary network to extract shape-related features, while a cost function based on these features is employed for the training of the model [38, 39]. Even though these approaches have led to improved results, it is not yet demonstrated whether the extracted shape-related features are optimal for the given task of segmentation. Therefore, a highly discriminative shape-based feature extraction scheme is proposed in this study for 3D anatomical organ segmentation, based on the extracted eigenvectors from the training dataset. To this end, a dedicated cost function is defined to take into account the underlying shape features in the form of eigenvectors to aid the network to avoid erroneous/unrealistic shape estimation.

The rest of this paper is organized as follows. In Section II, the datasets and the preprocessing are described; then, the shape-related feature extraction scheme and definition of the dedicated cost function will be elaborated. In Section III, the evaluation of the proposed cost function, concerning the same model with and without the dedicated loss function, will be described. Finally, the discussion and conclusion regarding the proposed approach will be covered in sections IV and V.

## II. MATERIALS AND METHOD

This section first describes the 3D medical image datasets, which were used in this study with different modalities (i.e., CT and MRI), and the preprocessing that was performed. Afterward, a comprehensive explanation is provided of the proposed shape-based cost function and the implementation details for further studies.

### A. Datasets and Preprocessing

Two datasets were exploited in this study to examine the efficiency of the proposed shape-based cost function in the training of the CNN models for 3D medical image segmentation. The first dataset belongs to the Decathlon medical segmentation challenge [40], which consists of 260 T1-weighted MR images of the hippocampus with dimensions of $(31 \sim 43) \times (40 \sim 59) \times (24 \sim 47)$ and 1 $mm^3$ voxel size. Manual delineation masks for the head and body of the hippocampus were available for each subject. Preprocessing steps for this dataset consisted of 1) zero-padding the entire images and corresponding masks to a size of $48 \times 64 \times 48$, 2) merging the labels for the hippocampus head and body to create a single binary mask for the whole hippocampus, 3) normalizing the intensity of the MR images to a fixed range of 0-1, and 4) randomly selecting 90% of the whole data for training (234 cases) and 10% (26 cases) for evaluation (external test dataset).

The second dataset, called SegTHOR [41], is a public dataset consisting of 40 thoracic CT scans with lung cancer or Hodgkin's lymphoma. Four organs were manually segmented (considered as ground-truth) for each subject, including Esophagus, Heart, Trachea, and Aorta. For this study, only the heart segmentation was considered from this dataset. To this end, heart labels and CT images were cropped in $160 \times 144 \times 64$ voxel size. The preprocessing steps carried out for this dataset could be summarized as follows: 1) the entire voxel spacings were resampled to (1,1,2) mm in the X, Y, and Z axes, respectively (since the images were collected from various CT scans); 2) the CT values within the range of -800, 400 HU were normalized to a 0~1 range; and 3) 80% of the whole dataset was randomly selected for the training (32 cases) and the leftover (8 cases) for testing the model.

### B. Shape-based loss function

Recent studies showed that neural networks may fail (or show suboptimal performance) to learn the shape and morphological features [18]. In this regard, final segmentation results may bear sorts of defects, such as unrealistic anatomy representation, islands of voxels, and patchy patterns. To address these issues and enhance the performance of the CNN models in capturing the underlying/valid shape variations, attempts were made to define a shape-based cost function dedicated to the target organ/anatomy. To this end, the underlying shape characteristics in the form of eigenvectors/eigenvalues generated by the PCA were exploited [42].

Figure 1 illustrates the entire procedure of the proposed method. The entire training masks were vectorized to constitute a shape-space or the probability of shape variations for the target organ and to generate the elements of the shape-space. Given the shape-space of the target organ, the shape-based cost function exploits the extracted eigenvectors from the PCA-based shape modeling to constrain/encourage the CNN model to generate/converge to a valid estimation of the target organ structure. The dedicated loss function then relies on the extracted eigenvectors (will be elaborated in the following) to disregard/penalize the estimated structures which are not consistent with the principal shape variations. The proposed cost function takes advantage of the weighted sum of a shape-based cost compartment and the Binary Cross-Entropy (BCE) cost element to evaluate the accuracy of the estimated structure.

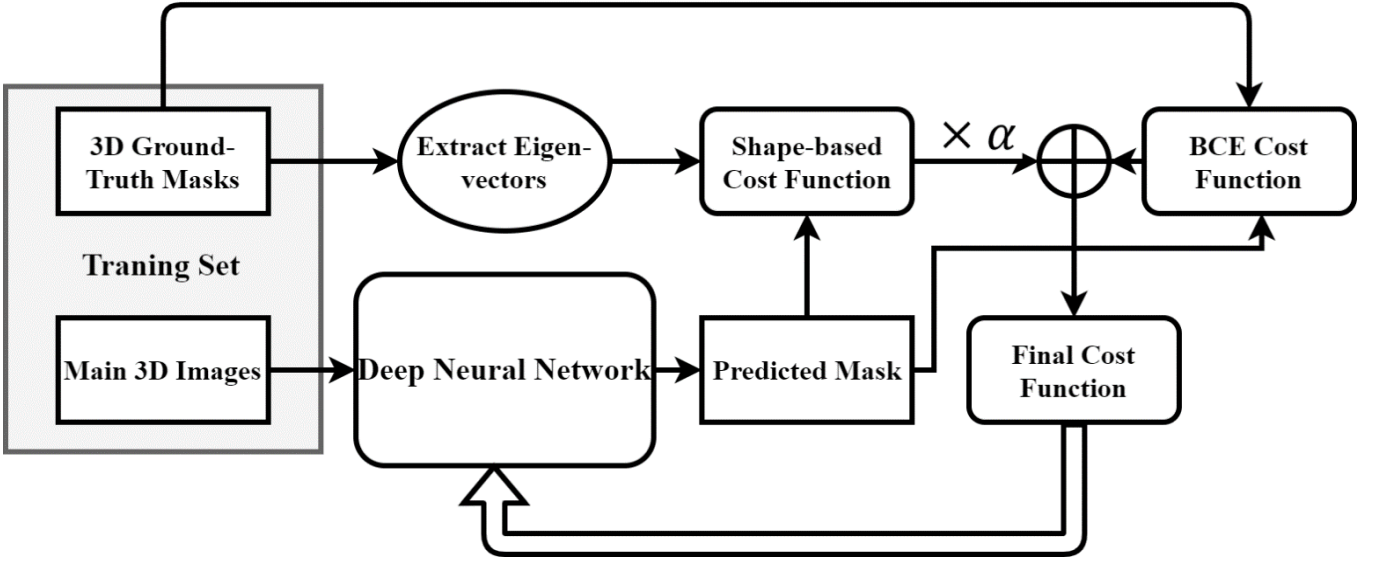

**Figure 1.** Block diagram of the proposed shape-based segmentation method. First, a shape-space is constructed using flattened 3D annotated masks of the training set to extract eigenvectors and eigenvalues using PCA. These extracted eigenvectors are used in the following stages to extract the reconstruction weights of the predicted labels and form shape-based cost functions. A weighted sum of the proposed cost function and the Binary Cross-Entropy was used to penalize/guide the estimation of the target structure in a voxel-wise manner.

The vectorized 3D labels are assumed as $L_1, L_2, ..., L_n$, where $n$ is the number of training volumes. For the extraction of the eigenvalues and eigenvectors, the following procedure is performed:

$$\boldsymbol{\theta}_i = \boldsymbol{L}_i - \boldsymbol{\mu}, \qquad \boldsymbol{\mu} = \frac{1}{n}\sum_{i=1}^{n} \boldsymbol{L}_i$$

$$C = \frac{1}{n}\sum_{i=1}^{n} \boldsymbol{\theta}_i \boldsymbol{\theta}_i^T = \frac{1}{n}\Theta\Theta^T, \qquad \Theta = [\boldsymbol{\theta}_1, \boldsymbol{\theta}_2, ..., \boldsymbol{\theta}_n]$$

$$C\boldsymbol{u}_i = \frac{1}{n}\Theta\Theta^T \boldsymbol{u}_i = \lambda_i \boldsymbol{u}_i \qquad (1)$$

Where $\boldsymbol{\mu}$, $C$, $\boldsymbol{u}_i$ and $\lambda_i$ denote the mean, covariance matrix, eigenvector, and eigenvalue of the target shape-space, respectively. In this study, a 3D label map of the heart had a dimension of 160×144×64 voxels, which were vectorized into an array with 1,474,560 elements. This vector led to a covariance matrix with a size of 1,474,560 by 1,474,560 elements, which is extremely large for normal computer memories (RAMs). A feasible approach to calculate the eigenvectors and eigenvalues of such a large covariance matrix would be the reformulation of Eq. 1 as

$$\frac{1}{n}\Theta^T\Theta v_i = \lambda_i v_i$$

$$\left(\frac{1}{n}\Theta\Theta^T\right)\Theta v_i = \lambda_i(\Theta v_i), \qquad u_i = \Theta v_i \qquad (2)$$

Wherein instead of calculating $\Theta\Theta^T$ using the same size of the covariance matrix, $\Theta^T\Theta$ is calculated, which has a dimension of $n\times n$ ($n$ is the number of training data). This matrix size would be manageable for computers with normal configurations. In this regard, the eigenvectors ($u_i$) defined in Eq. 1 could be obtained from Eq. 2 in a memory-efficient fashion [43]. This procedure was followed to extract the eigenvectors of hippocampus and heart label maps referred to as Eigen-organ (Eigen-Hipp and Eigen-heart, respectively). Figure 2 depicts the first two and the last reshaped extracted Eigen-organs in axial, coronal, and sagittal views.

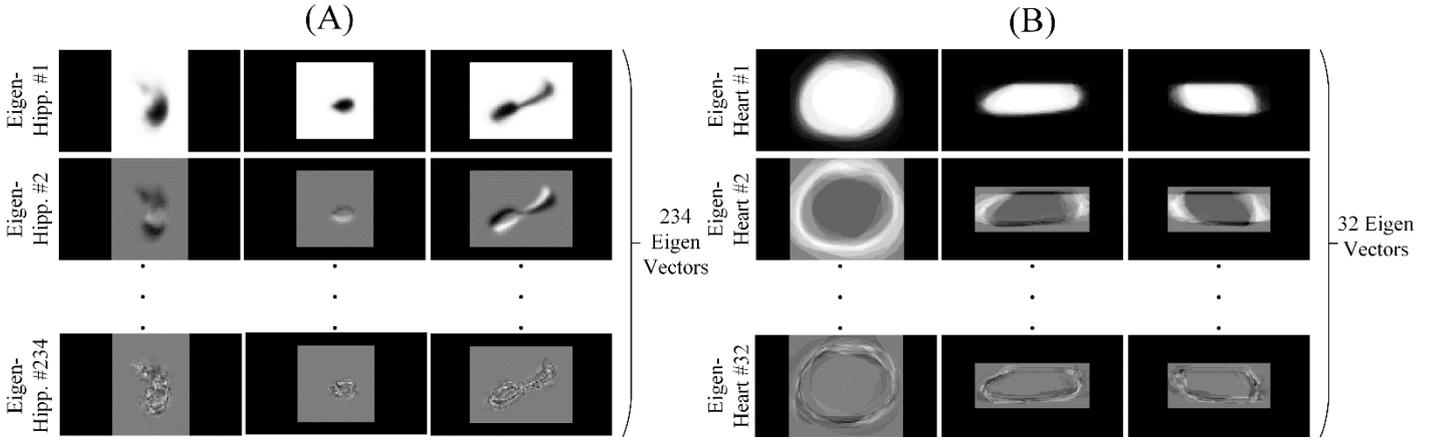

**Figure 2.** Extracted Eigen-organs for the A) hippocampus and B) heart in axial, coronal, and sagittal views.

Given the organs' eigenvectors and eigenvalues, the primary structures could be reconstructed using Eqs. 3 and 4, wherein $\widehat{L}_i$ is the reconstructed structure, and $w$ denotes the weights vector for reconstruction.

$$L_i \times U = w, \qquad U = [u_1, u_2, \ldots, u_n] \qquad (3)$$

$$\widehat{L}_i = U \times w^T \qquad (4)$$

The organ structures could be reconstructed using the primary components formulated in Eq. 3. In this light, attempts were made to exploit these primary reconstruction weights to discriminate valid and non-valid estimations of the target structure. To this end, a set of random and corrupted labels (equal to the number of subjects in the training sets) were generated (referred to as Non-Hipp and Non-Heart for the hippocampus and heart datasets, respectively), and the reconstruction weights for both ground-truth organs, as well as non-organ labels, were extracted. The histogram of the extracted weights for true and corrupted hippocampus and heart labels are presented in Figures 3.A and 3.C, respectively. These histograms demonstrated the discriminative ability of the reconstruction weights to distinguish between genuine organ shape and erroneous estimations.

In fact, each reconstruction weight is a vector consisting of $n$ elements ($n$ is equal to the number of subjects in the training dataset) which could be regarded as a probability distribution of specific shape reconstruction factors.

Regarding Figure 3, the histograms of ground-truth weights are different, wherein the kurtosis value of true organ and non-organ weights would be discriminative, as illustrated in Figures 3.B and 3.D for the hippocampus and heart, respectively.

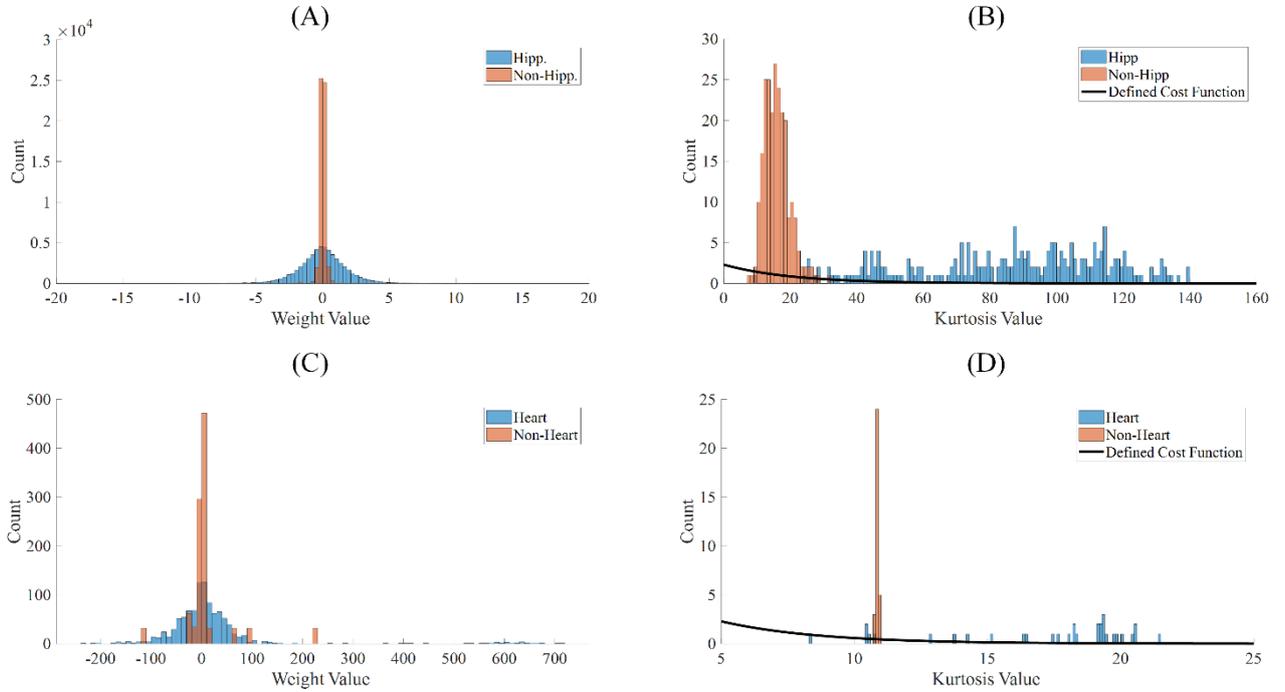

**Figure 3.** A) Histogram of the primary reconstruction weights for the true hippocampus and Non-Hipp labels. B) Kurtosis histogram of the hippocampus and Non-Hipp labels (the defined cost function is illustrated in black line). C) Histogram of the primary reconstruction weights for the true heart organ and Non-Heart labels D) Kurtosis histogram of the heart and Non-Heart labels (the defined cost function is illustrated in black line). It is evident that most of the kurtosis values of the actual organ labels are higher than non-organ labels; thus, a cost function was defined based on the kurtosis of reconstruction weights that penalize the network for lower kurtosis values.

As illustrated in Figure 3, the kurtosis of the primary reconstruction weights could serve as a discriminative factor for true organ and non-organ labels. In this light, a dedicated shape-based cost function was proposed, which exploits shape features in the form of primary reconstruction weights. Regarding the kurtosis values in Figures 3.B and 3.D, higher values for the reconstruction weights of the true organs were observed, compared to the non-organ labels. To define the cost function, a decaying exponential function was employed to generate a small loss for high kurtosis values and vice versa. The shape-based component of the cost function (SBCF) is defined in Eq. 5.

$$SBCF = \beta e^{-\frac{x-\gamma m}{\delta \sigma}} \tag{5}$$

Where, $x$ is the kurtosis value of the reconstruction weights for the estimated shape (calculated by Eq. 3), $m$ and $\sigma$ denote the mean and standard deviation of the training set kurtosis, and $\beta$, $\gamma$, as well as $\delta$, are the tuning coefficients of the decay exponential curve. The final cost function is a weighted sum of the SBCF and BCE cost elements (Eqs. 6 and 7).

$$BCE = -\frac{1}{N}\sum_{i=1}^{N} y_i \cdot \log(p_i) + (1 - y_i) \cdot \log(1 - p_i) \tag{6}$$

$$proposed\ loss = \alpha SBCF + BCE \tag{7}$$

Here, $y_i$ and $p_i$ are the references and predicted labels, respectively. In addition, $N$ denotes the number of pixels in the output image, and $\alpha$ is the weight factor.

*C. Implementation Details*

To validate the effectiveness of the proposed loss function, a 3D fully convolutional neural network (FCN) was developed using dilated convolutional kernels to achieve larger receptive fields with no extra trainable parameters. Figure 4 illustrates the network architecture consisting of 20 convolutional layers, wherein the first seven layers, with 3×3×3 convolutional kernels and without dilation, were exploited to extract low-level features. In the following layers, dilation factors of 2 and 4 were applied to the convolutional kernels to detect medium- and high-level features, respectively. Residual connections were established to link every two blocks. Moreover, each layer included batch normalization and rectified linear unit as activation function, except for the last layer where a sigmoid was used for the final segmentation output [44].

The FCN was trained two times: first, with the BCE loss function, and the second time, with the proposed loss function formulated in Eq. 7. The Adam algorithm was utilized for the optimization of the models and a base learning rate of 0.01 with a decay factor of 0.1.

To train the FCN architecture with the proposed cost function, the hyper-parameters of $\beta$, $\gamma$, and $\delta$ (Eq. 5) were set to 0.04, 1, and 0.7 for the hippocampus dataset, and 0.5, 0.6, and 1 for the heart dataset, respectively. It should be noted that hyper-parameter setting was achieved based on the histograms in Figures 3.B and 3.D to have high values in low kurtosis and vice versa. The grid search was followed for the optimization of the hyper-parameters $\alpha$ (Eq. 7), wherein the best results were observed with $\alpha = 0.1$ and $\alpha = 2$ for the hippocampus and heart datasets, respectively. The training of the models was performed on a Tesla P100-PCIE-16GB GPU, in which the number of 3D convolutional kernels at each layer (shown by $K$ in Figure 4) was set to 16 for the hippocampus dataset and 3 for the heart dataset to achieve the maximum size of feature maps regarding the RAM capacity.

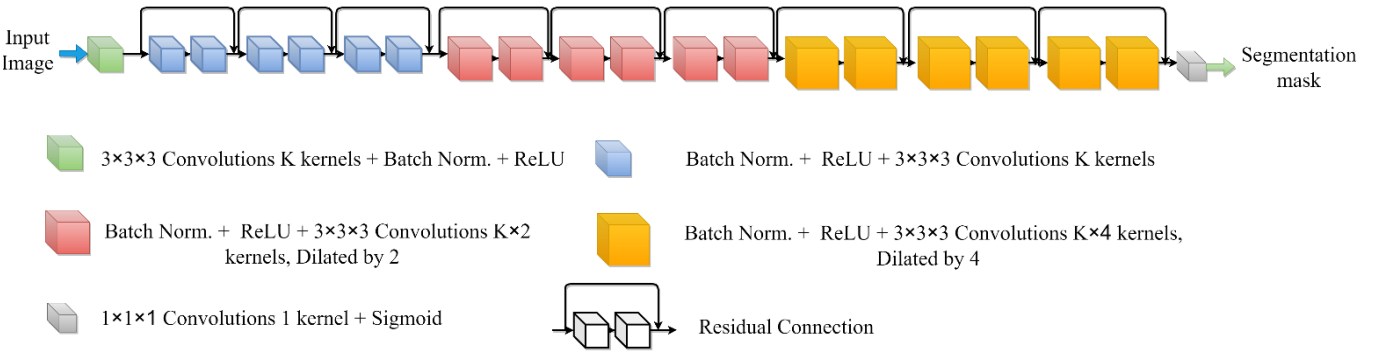

**Figure 4.** 3D convolutional network architecture used to evaluate the proposed shape-based cost function. This architecture exploits dilated convolution in deeper layers to expand the receptive field of the filters and residual connections to prevent gradient vanishing.

## D. Evaluation metrics

Standard evaluation metrics, including Dice score (Eq. 8), Hausdorff distance (Eq. 9), and Jaccard index (Eq. 10), were employed to assess the outcomes of the segmentation networks.

$$Dice\ Score = \frac{2|X \cap Y|}{|X| + |Y|} \quad (8)$$

$$d_H(X,Y) = max\{d_{XY}, d_{YX}\}$$
$$d_{XY} = \max_{x \in X} \min_{y \in Y} d(x,y) \quad (9)$$

$$J(X,Y) = \frac{|X \cap Y|}{|X \cup Y|} \quad (10)$$

Here, X is the network output, Y is the ground-truth, and $d_{XY}$ is the maximum of the minimum distances of contour X from Y.

## III. RESULTS

To evaluate the effectiveness of the proposed shape-based loss function, the same networks with an identical structure were trained twice with different cost functions, including the shape-based and BCE loss functions. Table 1 summarizes the calculated evaluation metrics for the two datasets obtained from the two deep learning models trained with different loss functions. Evidently, the proposed loss function led to a significant improvement in segmentation accuracy regarding the entire quantitative metrics. To statistically examine the significance of differences between the two deep learning models, a paired t-test analysis was performed where P-values smaller than 0.05 were considered statistically significant. The third row for each dataset in Table 1 presents the calculated P-value for the segmentation metrics, which implies significant improvement achieved by the proposed shape-based cost function.

**Table 1.** Segmentation metrics comparing the performance of the shape-based cost function with the Binary Cross-Entropy cost function for the hippocampus and heart segmentation.

| Dataset | Method | Metric | | |
| --- | --- | --- | --- | --- |
| | | Dice | Jaccard | Hausdorff distance |
| Hippocampus segmentation | BCE cost | 0.81±0.03 | 0.69±0.05 | 3.10±0.44 |
| | SBCF (ours) | **0.86±0.03** | **0.75±0.05** | **2.58±0.35** |
| | P-value | 0.000006 | 0.000003 | 0.000007 |
| Heart segmentation | BCE-loss | 0.74±0.07 | 0.59±0.08 | 9.31±1.70 |
| | SBCF (ours) | **0.87±0.05** | **0.77±0.08** | **8.60±1.39** |
| | P-value | 0.0032 | 0.0026 | 0.33 |

Figures 5 and 6 illustrate the representative segmentations of the hippocampus and heart, respectively, by the two deep learning models in axial, coronal, sagittal, and 3D views. The first rows in both figures depict the ground-truth labels annotated by the experts; the second and third rows show the outputs of the deep learning models trained with BCE and shape-based cost functions, respectively. Visual inspection of the resulting organ masks revealed that defects, such as fragmented structures, islands of voxels, and topological inconsistencies, did not occur when the shape-based cost function was implemented (besides the overall improvement in the segmentation accuracy). Figure 5, depicting the magnified hippocampus masks, demonstrates that the BCE cost function may lead to an island of voxels segmented erroneously (fragmentation of structure in the hippocampus shape). This issue was resolved when the shape-based cost function was employed for the network training, and higher accuracy, as well as shape consistency, was achieved. Similarly, Figure 6 shows a shape inconsistency in the heart segmentation brought by the BCE loss function. The trained network with the BCE cost function predicted a mask that does not contain the entire heart structure (a shrunk version of the heart mask is predicted); however, the shape-based model resulted in a valid estimation of the heart structure.

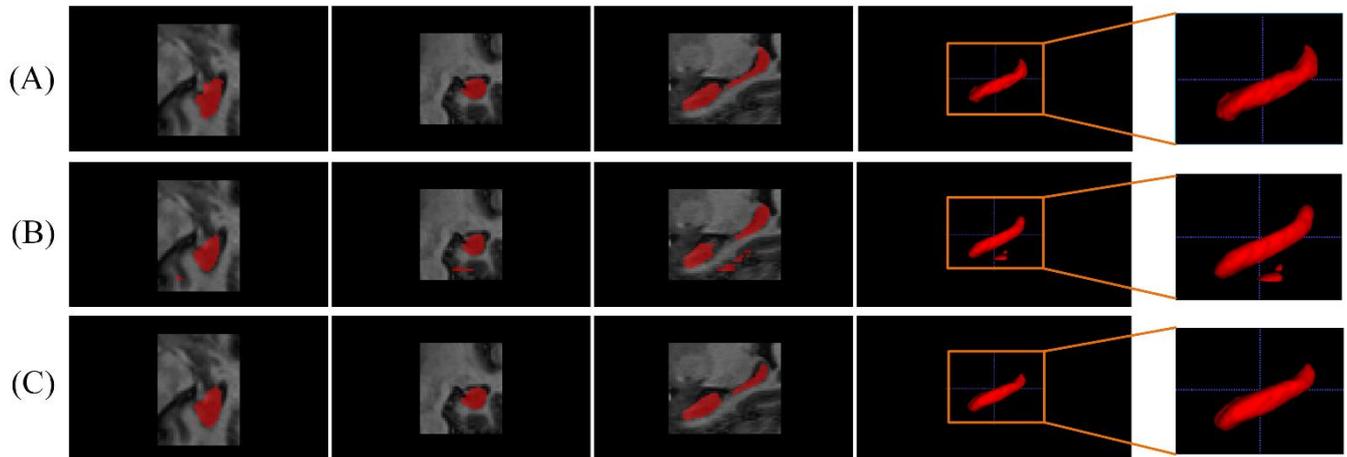

**Figure 5.** Hippocampus segmentation: A) ground-truth mask, B) prediction by the deep learning model trained with the BCE cost function, and C) prediction by the deep learning model trained with the shape-based cost function in axial, coronal, sagittal, and 3D views from left to right. Fragmentation of the hippocampus structure is evident in the magnified prediction of the network trained with a BCE cost function. This inconsistency has been solved using the proposed shape-based cost function.

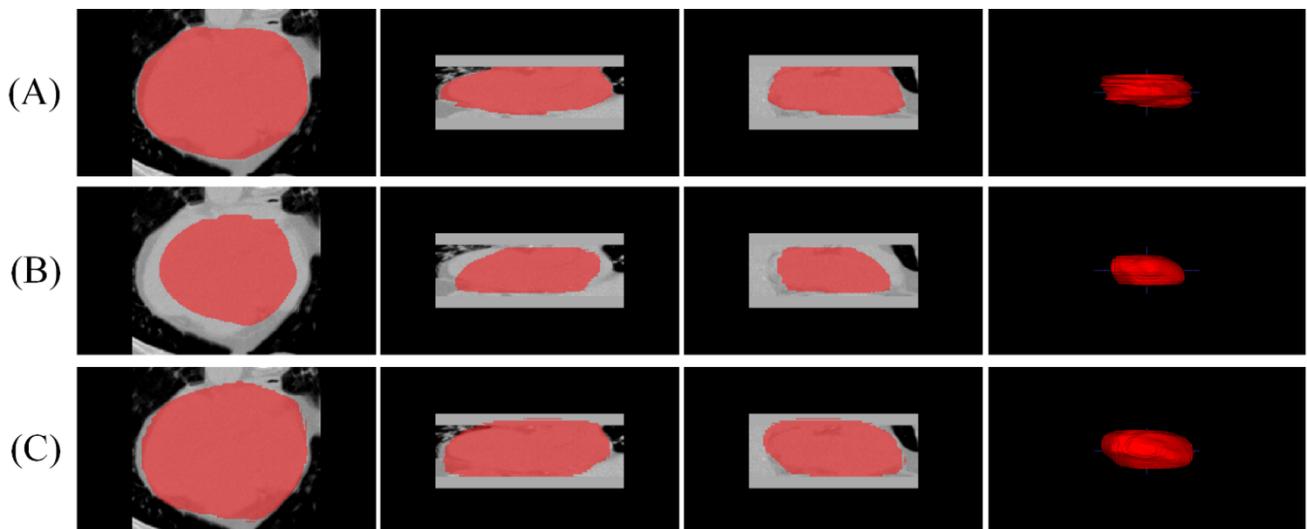

**Figure 6.** Heart segmentation: A) ground-truth mask, B) prediction by the deep learning model trained with the BCE cost function, and C) prediction by the deep learning model trained with the shape-based cost function in axial, coronal, sagittal, and 3D views from left to right.

Figures 7.A and 7.B illustrate the box plots of the quantitative metrics, including Dice score, Jaccard index, and Hausdorff distance for the two cost functions, the hippocampus and heart segmentations, respectively. For better visualization, Hausdorff distance boxplot values were normalized to their maximum values of 4.24 and 11.66 for the hippocampus and heart, respectively.

As shown in Table 1 and Figure 7, exploiting the shape-based cost function resulted in improved mean performance accuracy, as well as a reduced standard deviation, across the test dataset. Regarding Figure 7.A, there is an outlier for the hippocampus segmentation by the shape-based loss function. This case is presented in Figure 8, wherein the ground-truth mask is indicated in green color, and BCE, as well as shape-based loss function predictions, are indicated in red. The dice score of this case for the shape-based cost function is 0.73, and for the BCE cost function is 0.77. Visual inception did not reveal many differences between these two predictions; however, the magnified versions show an unwanted island of voxels in the BCE prediction.

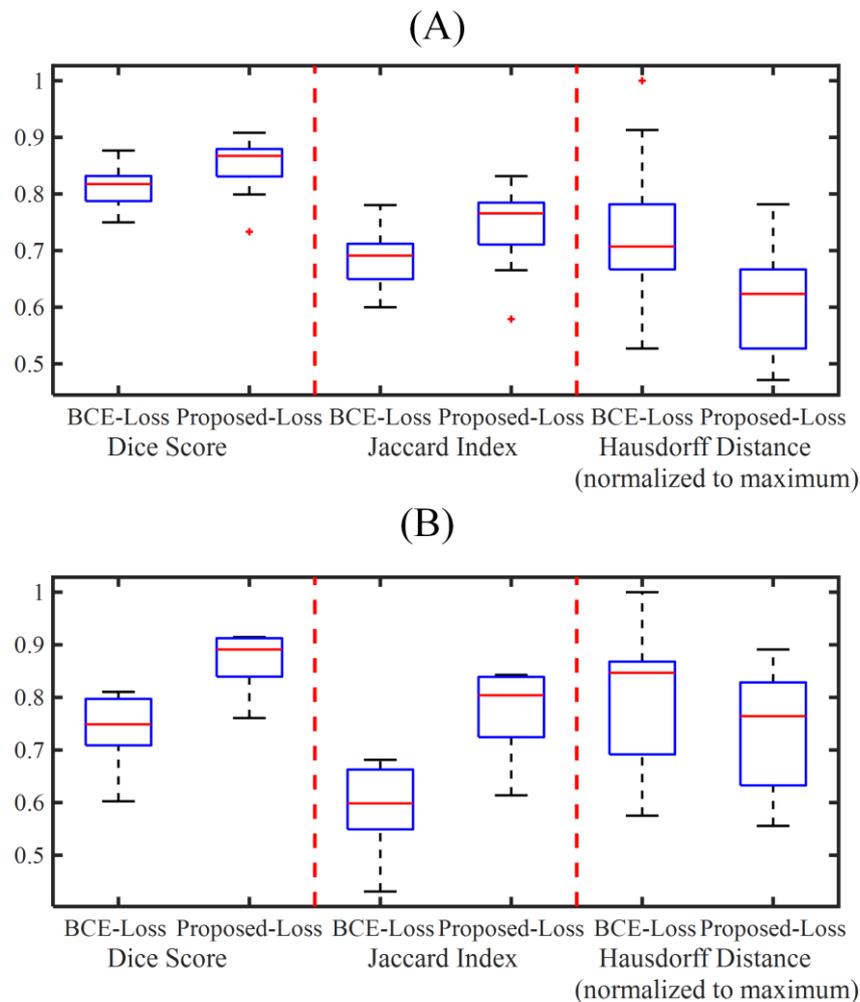

**Figure 7.** Boxplots of the quantitative metrics, including Dice score, Jaccard index, and Hausdorff distance, for A) hippocampus and B) heart segmentations.

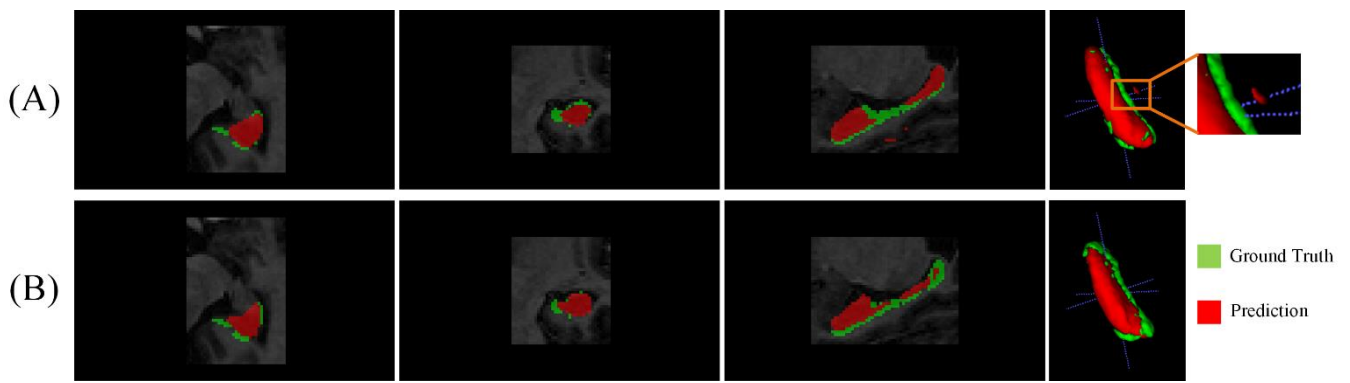

**Figure 8.** The outlier in the hippocampus segmentation produced by the shape-based loss function: A) ground-truth (in green) and BCE cost function output (in red) and B) ground-truth (in green) and shape-based cost function output (in red). Although the Dice score for the BCE cost function is higher, this loss function led to an erroneous island of voxels.

Figure 9 shows the histogram of kurtosis values obtained from analyzing the ground-truth, prediction by the BCE cost function, and prediction by the proposed shape-based cost function in blue, green, and orange, respectively, for the hippocampus and heart segmentations. The defined shape-based cost function, indicated by the black curve, forces/encourages the deep learning model to achieve higher kurtosis values, which leads to a realistic shape prediction with higher segmentation accuracy.

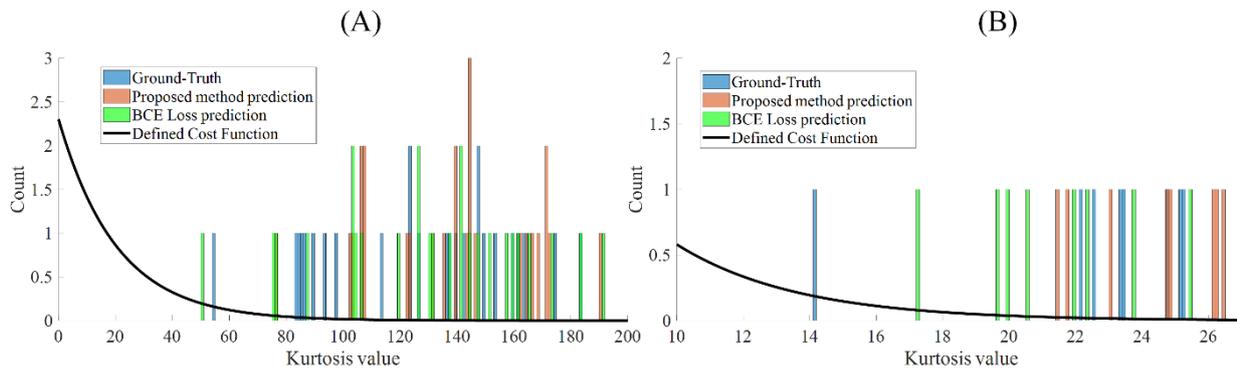

**Figure 9.** Histogram of the kurtosis values obtained from analyzing the ground-truth (blue), prediction by the BCE cost function (green), prediction by the proposed shape-based cost function (orange) for A) hippocampus B) heart in the external test dataset. In most cases, a higher kurtosis value led to a realistic shape, as promoted by the shape-based cost function.

## IV. DISCUSSION

Medical image segmentation plays a critical role in diagnosis and treatment planning procedures in clinical practices. Therefore, it is necessary to propose semi-automated or automated methods to speed up this process and prevent human errors. Deep learning-based segmentation methods have achieved tremendous results in medical image segmentation, in comparison with earlier/conventional approaches. Despite improvements in medical image segmentation brought by deep learning-based approaches, yet there are sorts of shortcomings associated with these approaches. Deep CNNs, as a case in point, may exhibit the suboptimal capacity to learn the shape and morphological features of the target organ/structure for segmentation [18], which would lead to the fragmentation of integrated structures, islands of voxels, and holes in the final network outcome. To address these defects, many approaches have been proposed, including post-processing the network outputs or providing the network with prior knowledge regarding the underlying shape variations of the target structures to improve prediction accuracy [22, 29, 37]. Nevertheless, the issue of shape and morphology consistency is not resolved in deep learning approaches.

An efficient solution to address this issue in deep learning-based segmentation frameworks is to define a cost function that constrains the shape estimation space to result in a valid/reasonable shape. In this regard, efforts have been made to guide convolutional networks to predict valid/realistic shapes through a pre-trained autoencoder compartment that encodes the ground-truth and predicted mask to a lower dimension vector and uses these vectors to define a cost function [38, 39]. The reputation of deep neural networks as a black box indicates the lack of transparency or interpretability of how exactly the input data is transformed into model outputs [45]. In this light, the exact role of the proposed pre-trained autoencoder on the shape consistency of the estimated structures is not evident. This issue could be addressed by directly penalizing the output shape by a shape-based cost function rather than using features extracted with an encoder.

The cost function proposed in this study relies on the underlying features obtained from the PCA to prevent the deep neural network from generating erroneous/unrealistic estimations. The key idea is that in human organ segmentation, there are relatively small discrepancies across different patients (compared to the natural images). If a shape-space is created from the target organs, invalid shape estimations could be strongly penalized since they are not consistent with the underlying properties of the shape-space. A merit of this approach is that the shape-space is not generated necessarily from the training dataset and any dataset related to the target structure could be exploited. To distinguish between a member and a non-member to the shape-space, eigenvectors and eigenvalues were extracted, wherein a reconstruction of the target shape would be feasible using a weighted sum of eigenvectors. The proposed cost function tends to discriminate/contrast between a realistic/valid shape and an unrealistic/invalid shape through assessing the reconstruction weights of the network prediction. The initial experiments in the present study demonstrated that the kurtosis of the reconstruction weights is more effective in discriminating between realistic and unrealistic shapes (higher values of kurtosis will lead to a more realistic shape). The proposed cost function not only improved the overall accuracy of the segmentation models, but also realistic/valid shapes have been predicted without any islands of voxels or shape inconsistencies (no outliers were observed in the outcomes of the proposed model).

As mentioned before, a great merit of the proposed framework is that any dataset related to the target organ could be exploited to create the shape-space, and the training dataset should not necessarily contain a specific number of subjects. For example, in the segmentation of the heart from CT images, the available masks from the MR segmentation could be exploited for the generation of the shape-space. Nevertheless, the proposed cost function for the segmentation task suffers from the fact that this cost function could be employed for the target structures with predictable shapes. The performance of the proposed cost function might be limited for lesion or pathology segmentation which may not follow specific shape variation [46, 47]. However, this issue warrants a separate study to investigate the impact of using shape prior in lesion and pathology segmentation.

This study presented a confirmation of the concept regarding the application of the PCA-derived cost function in human organ segmentation. For future studies, the researchers would propose a shape-based cost function for multi-organ segmentation and the utilization of an effective alternative to the kurtosis of reconstruction weights, such as a neural network, to differentiate between valid and invalid estimated shapes.

## V. CONCLUSION

In this study, a novel shape-based cost function was proposed for the deep learning segmentation models to reinforce realistic/valid shape estimation from medical images. This cost function was defined dedicatedly for the target organ/anatomy using the PCA. The dedicated shape-based cost function not only led to the overall improvement of the segmentation accuracy, compared to the conventional cost functions, but also resulted in no outliers or estimated structures with noticeable defects, such as holes, patchy voxels, and unrealistic shapes.